\documentclass[12pt,preprint]{aastex}

\begin{document}

\title{The Cause of Photospheric and Helioseismic Responses
to Solar Flares: High-Energy Electrons or Protons?}
\author{A. G. Kosovichev}
\affil{W.W.Hansen Experimental Physics Laboratory, Stanford
University, Stanford, CA 94305, USA}

\begin{abstract}
Analysis of the hydrodynamic and helioseismic effects in the photosphere
during the solar flare of July 23, 2002, observed by Michelson Doppler
Imager (MDI) on SOHO, and high-energy images from RHESSI
shows that these effects are closely associated with sources
of the hard X-ray emission, and that there are no such effects
in the centroid region of the flare gamma-ray emission.
These results demonstrate that contrary to
expectations the hydrodynamic and helioseismic responses
(''sunquakes") are more likely to be caused by accelerated electrons than by
high-energy protons. A series of multiple impulses of high-energy
electrons forms a hydrodynamic source moving in the photosphere with
a supersonic speed. The moving source plays a critical role in the
formation of the anisotropic wave front of sunquakes.
\end{abstract}
\section{Introduction}
``Sunquakes", the helioseismic response to solar flares, are caused
by strong localized hydrodynamic impacts in the photosphere during
the flare impulsive phase. The helioseismic waves are observed
directly as expanding circular-shaped ripples in SOHO/MDI
Dopplergrams, which can be detected in Dopplergram movies and as a
characteristic ridge in time-distance diagrams,
\citep{Kosovichev1998, Kosovichev2006a}, or indirectly by
calculating integrated acoustic emission \citep{Donea1999,
Donea2005}. Solar flares are sources of high-temperature plasma and
strong hydrodynamic motions in the solar atmosphere. Perhaps, in all
flares such perturbations generate acoustic waves traveling through
the interior. However, only in some flares the impact is
sufficiently localized and strong to produce the seismic waves with
the amplitude above the convection noise level. It has been
established in the initial July 9, 1996, flare observations
\citep{Kosovichev1998} that the hydrodynamic impact follows
the hard X-ray flux impulse, and hence, the
impact of high-energy electrons. Nevertheless, a common paradigm is
that the sunquake events are caused by accelerated protons because
protons carry more momentum and penetrate deeper into the solar
atmosphere than electrons, which loose most of their energy in the
upper chromosphere. This paradigm is not easy to test because
the gamma-ray emission, which indicates the presence of high-energy
protons, is rarely observed.

In a large X17 flare of October 28, 2003, the gamma-ray emission
observed by RHESSI was located close to the hard X-ray sources and
two of the three places of the phototospheric impacts (sunquake
sources) \citep{Kosovichev2006a}. Because of the close locations of
the hard X-ray and gamma-ray sources these observations could not
exclude the possibility of the proton or mixed electron-proton
 impacts \citep{Zharkova2007}.

However, in one event, X4.8 flare of July 23, 2002, the hard X-ray
and gamma-ray sources were significantly separated from each other.
The centroid of the $\gamma$-ray 2.233 MeV neutron-capture
emission was found to be displaced by $20"\pm 6"$
(with 5-sigma confidence) from that of the
$0.3-0.5$ MeV X-ray emission implying a difference in acceleration and/or
propagation between the accelerated electrons and ions
\citep{Hurford2003}. Therefore, this flare provides a unique
opportunity to investigate the photospheric and helioseismic
responses separately for high-energy electrons and protons. In this
Letter, I present results of the analysis of the relationship
between the hard X-ray and gamma-ray emissions and the hydrodynamic
and seismic signals in the photosphere, using data from RHESSI
\citep{Lin2002} and MDI on SOHO \citep{Scherrer1995}.
RHESSI provides X-ray/gamma-ray imaging spectroscopy from 3 keV to
17 MeV with angular resolution $2.3''-3'$ ($35''$ at gamma-ray energies)
 over the full Sun.  MDI
measures the Doppler velocity and the line-of-sight magnetic field of
the photospheric plasma every minute with 2 arcsec/pixel resolution
also over the full Sun.

\section{Analysis of SOHO/MDI and RHESSI data}

\begin{figure}
\centerline{\includegraphics[scale=0.8]{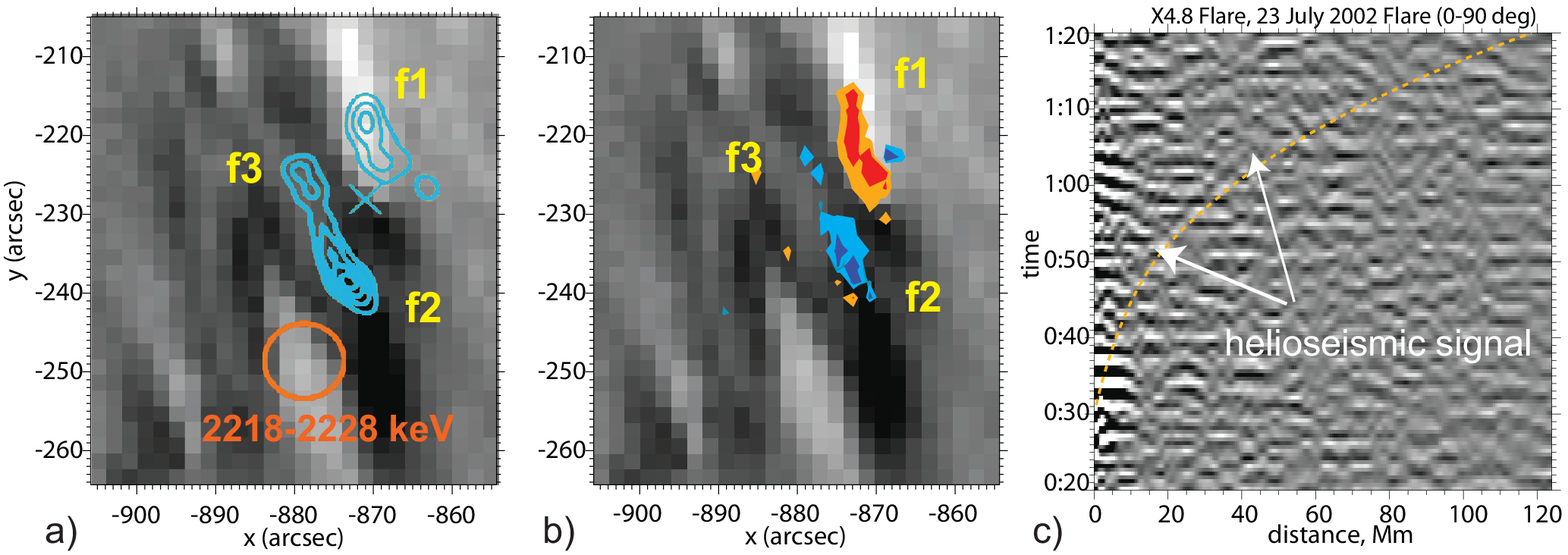}} \caption{a) RHESSI
observations of HXR (f1, f2, f3) sources (blue contours show the 50--100 keV map
with $3''$ resolution, and the blue cross shows the HXR centroid)
and a gamma-ray source (orange circle shows the location of the
gamma-ray centroid with $1\sigma$ error) (from \citet{Krucker2003}). The
gray-scale background is MDI magnetogram ($\pm 600$ G range). b)
MDI observations of Doppler velocity sources. Orange contour lines
show positive (red-shift) velocity greater than 0.6 km/s; the red
contours show 1 km/s; the blue contour lines show negative
(blue-shift) velocity  of -0.6 km/s, and dark blue shows -0.7 km/s.
c) Time-distance map revealing a seismic wave front, which travels
in the North-West direction from the location of source f1.
The yellow dashed curve is a theoretical time-distance relation
for helioseismic acoustic waves.} \label{fig1}
\end{figure}

X4.7 flare of July 23, 2002, was the first flare, for which
gamma-ray images were obtained \citep{Hurford2003}. Other examples of
flares with RHESSI gamma-ray images are given by \citet{Hurford2006}.
The properties
of the gamma-ray and hard X-ray emissions, and also other aspects of
the July 23, 2002, flare, are discussed in the RHESSI special issue of ApJ Letters
(v.595,no.2, 2003). The RHESSI observations revealed three hard
X-ray sources and a gamma-ray source. Their positions in the MDI
magnetogram are shown in Figure~\ref{fig1}a \citep{Krucker2003}. The
hard X-ray sources (marked as f1, f2 and f3) were located on both
sides of the magnetic neutral line. The morphology of the
gamma-ray emission was not resolved but
it could not have been more than 1 arcminute (FWHM) in extent and its centroid
was $20''\pm 6$ south from the centroid of the hard X-ray sources and about $30''$
from the f1 source.

The MDI Dopplergrams  show strong impulsive variations close to the hard
X-ray sources, but no impulsive variations in the region of the
gamma-ray source centroid, or anywhere outside the hard X-ray sources.
Figure 1b shows the positions of the impulsive
Doppler velocity signals in the photosphere during the flare impulsive phase
between 00:27 UT and 00:36 UT, July 23, 2002.
This flare was rather close to the limb (coordinates of the flare sources
are given in Fig.~\ref{fig1}); the distance from the disk
center was approximately 70 degrees. Thus, it is possible that the
projection effect contributed to the opposite sign of the Doppler
shift across the neutral line, if the angle between the magnetic field
lines directing the plasma motion at the footpoints  (Fig.~\ref{fig2})
and the line of sight changes the sign. However, there might be other
reasons related to flare hydrodynamics, which should be explored.

The strongest Doppler signal, corresponding to a
downward plasma motion, appeared near the X-ray source, f1. Its position
moved during the impulsive phase in the North direction. This motion
is discussed in more detail in the next section. The time dependence
of the velocity signal at source f1 corresponds very well
(with the correlation coefficient of 0.8) to
the total hard X-ray flux in the 50-300 keV range (Fig.~\ref{fig3}c,d).
The gamma-ray emission (Fig.~\ref{fig3}e) is delayed by $\sim 100$ sec
because of the time for the neutrons to thermalize \citep{Murphy2003}.

The helioseismic waves are best visible at frequencies of about 5--6
mHz. To search for these waves the Dopplergrams were remapped into
the heliographic coordinates, tracked to
remove the displacement caused by the solar rotation, and then filtered
using a bandpass filter centered at 5.5 mHz with a FWHM of 2 mHz.
Then, the filtered Dopplergrams were remapped into the polar
coordinates, centered at various points including all hard X-ray
sources and the gamma source centroid region, and averaged azimuthally in several
angular sectors. The averaged signals are plotted as a function of
the radial distance and time, constituting time-distance propagation
diagrams. The diagrams were inspected for an elongated
characteristic ridge-like structure, which is caused by helioseismic
waves as predicted by the theory \citep{Kosovichev1995}, and
observed in other sunquake events \citep{Kosovichev1998,
Kosovichev2006a}. In this case, a rather weak ridge appeared only in
the propagation diagram, which was centered in the region of the
strongest impulsive Doppler signal at the f1 source and averaged in
the North-West quadrant. It can be identified in Fig. 1c for
distances between 20 and 40 Mm, just above the theoretical
time-distance relation for helioseismic
acoustic waves. The traveling wave front can be seen in the movie
of the frequency-filtered Dopplergrams. The observed signal is rather weak
 because a flare generates high-frequency acoustic waves, in which
 the plasma velocity is predominantly vertical \citep{Kosovichev1995}, and
 its line-of-sight projection is reduced by almost
2/3 due to the close-to-limb location. The amplitude of the line-of-sight
plasma velocity in this wave was about 20 m/s.
For the other central positions (including X-ray sources f2 and f3, and
X-ray and gamma-ray centroids) and sectors, the seismic waves were not detected.
This analysis puts the source of the seismic wave within
the lower red part of the Doppler source f1 in Fig.~\ref{fig1}b.
The start time estimated from the theoretical time-distance relation
in the ray approximation is 00:28--00:30 UT. It is interesting that
X-ray source f2 is marginally stronger than source f1, but the impulsive
Doppler signal is stronger in the f1 position.


The close correlation of the hydrodynamic and helioseismic responses
with the hard X-ray flux source and the absence of any significant
photospheric signal in the region of the gamma-ray centroid provide
 evidence that the source of the helioseismic waves is
associated with the high-energy electrons and not with the high-energy
protons.
We note that while the RHESSI data do not exclude the presence of
protons in the footpoints of the hard X-ray sources, for this conclusion
it is important that the proton flux in the gamma-ray centroid area
was not weaker than that in the HRX sources. This assumption is supported by
the RHESSI data. For further studies, it would be important to put
precise limits on the proton flux at the X-ray footpoints and
estimate the relative energetics of proton and electrons from RHESSI data.

\section{Moving hard X-ray and sunquake sources}

A characteristic feature of the seismic response in this flare and
several others \citep{Kosovichev2006a, Kosovichev2006b} is
anisotropy of the wave front: the observed wave amplitude is much stronger in
one direction than in the others. In particular, the seismic waves
excited during the October 28, 2003, flare had the greatest
amplitude in the direction of the expanding flare ribbons. The wave
anisotropy was attributed to the moving source of the hydrodynamic
impact, which is located in the flare ribbons
\citep{Kosovichev2006a, Kosovichev2006c}. The motion of flare
ribbons is often interpreted as a result of the magnetic
reconnection processes in the corona. When the reconnection region
moves up it involves higher magnetic loops, the footpoints of which
are further apart. Of course, there might be other reasons for the
anisotropy of the wave front, such as inhomogeneities in
temperature, magnetic field, and plasma flows. However, the source motion
seems to be quite important.

It is interesting that in the case of the July 23, 2002, flare the
seismic source identified in MDI Dopplergrams as a place of strong
Doppler shift in region f1 was moving mostly along the flare ribbon, and
consequently the seismic wave had the strongest amplitude in the
direction close to the direction of the source motion (but not precisely;
in this case, in addition to the other factors,
stronger foreshortening on the East side might have contributed
to the signal loss in the NE quarter).
The Doppler source motion nicely corresponds to the motion of the hard
X-ray source discovered by
\citet{Krucker2003}. Figure~\ref{fig4}a shows
the evolution of the hard X-ray sources, f1, f2 and f3, positions on
the magnetogram; and Fig.~\ref{fig4}b shows propagation diagrams
for this sources determined by \citet{Krucker2003}. From the top
panel of Fig.~\ref{fig4}b, the hard X-ray source, f1, traveled
approximately 7 Mm in 5 min; this corresponds to the mean speed of
approximately 20--25 km/s. The maximum speed according to
\citet{Krucker2003} reached 50 km/s.

Using the MDI Dopplergrams,  a similar time-distance propagation diagram was
constructed for the plasma photospheric velocity along the line of
motion of source f1. Figure \ref{fig4}c shows the Doppler velocity
along a 2-pixel wide strip along this line. This diagram shows that
the evolution of  the hydrodynamic impact source is very similar to
the evolution of the hard X-ray source (top panel in
Fig.~\ref{fig4}c). The mean speed of the hydrodynamic source was
also about 20-25 km/s.

Therefore, we conclude that the seismic wave was generated not by a
single impulse as was suggested in the sunquakes models of
\cite{Kosovichev1995, Medrek2000, Podesta2005} but by a series of
impulses, which produce the hydrodynamic source moving on the solar
surface with a supersonic speed. The seismic effect of the moving
source can be easily calculated by convolving the wave Green's
function (the wave signal from a point $\delta$-function type
source), $G(x-x_s,y-y_s,t)$ with a moving source function,
$S(x_s-V_xt,y_s-V_yt,t)$. The results of these calculations are
illustrated in Fig.~\ref{fig5}, which shows the wave front for a
source moving along the $x$-axis with a speed of 25 km/s. The strength
of this source varied with time as a Gaussian with FWHM of 3 min
(it is shown by black diamonds). The Green's function was calculated
by using the standard mode summation method \citep{Kosovichev1995}.
The strong anisotropy of the seismic wave is evident. Curiously,
this effect is quite similar to the anisotropy of seismic waves on
Earth, when the earthquake rupture moves along the fault
\cite[e.g.][]{Ben-Menahem1962}. Thus, taking into account the
effects of multiple impulses of accelerated electrons and moving
source is very important for sunquake theories. These effects will
be discussed in more detail in our future publications.

\section{Discussion}

The analysis of RHESSI X-ray and gamma-ray images and SOHO/MDI
Dopplergrams of the July 23, 2002, X4.8 solar flare revealed that
the hydrodynamic and seismic responses are closely associated the
hard X-ray emission, both spatially and temporally, but showed no
significant responses in the gamma-source centroid area. Because this flare
was one of strongest gamma-flares, and the hard X-ray and gamma-ray
sources were separated, these observations show that
the accelerated protons are unlikely to be a source of the hydrodynamic
response and sunquakes.
Furthermore, the detailed analysis of the dynamics of sunquake
sources in this Letter and in the paper by \citet{Kosovichev2006a}
reveals their close association with expanding flare ribbons and
rapid HXR source motion along the ribbons, and, thus, with the
magnetic reconnection process. The fast motion of these sources
results in strong anisotropy of the seismic waves, clearly observed
in the MDI data.

\begin{figure}
\centerline{\includegraphics[scale=0.5]{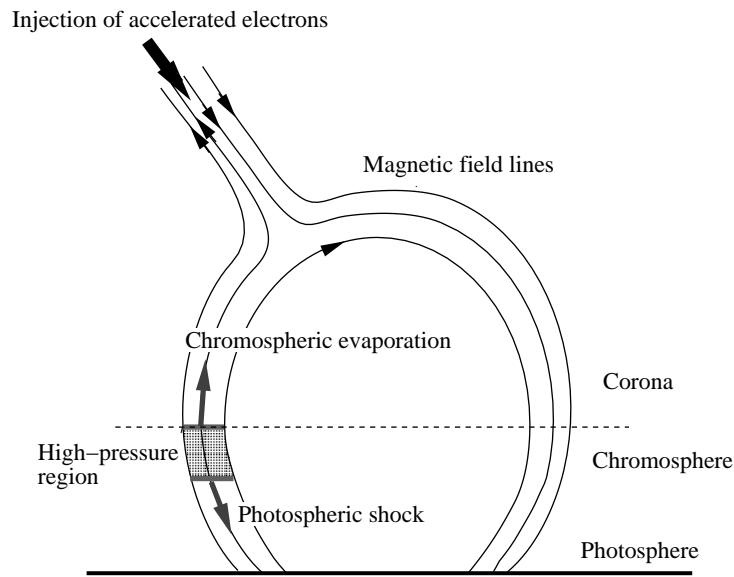}}
\caption{Illustration of the hydrodynamic thick-target model and the
mechanism of sunquakes. High-energy electrons accelerated in the
upper corona are injected along magnetic field lines into the
atmosphere, generate a hard X-ray emission in the loop footpoints
and heat the upper chromosphere to high temperature, producing a
high-pressure region. The high-pressure region expands producing
upward and downward propagating shocks. The downward shock reaches
the photosphere and causes a sunquake.
 }
\label{fig2}
\end{figure}

\begin{figure}
\centerline{\includegraphics[scale=0.5]{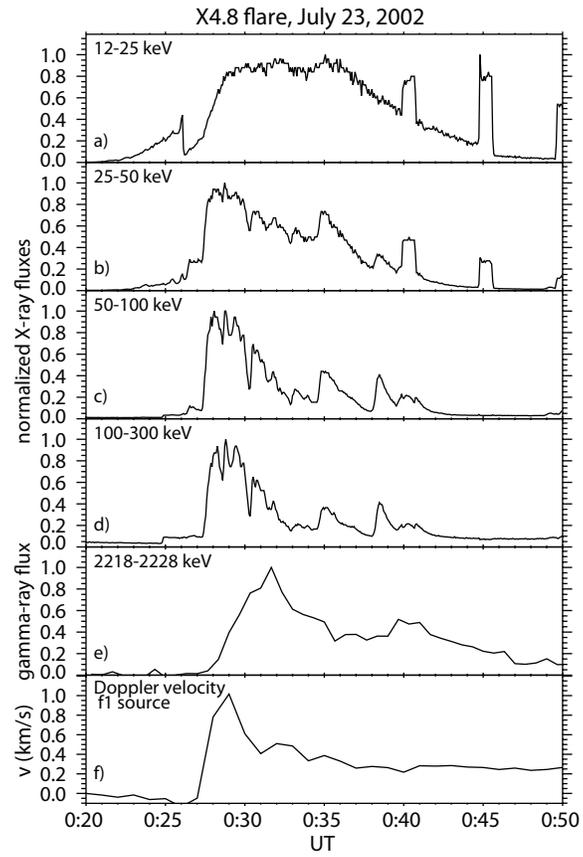}} \caption{Integrated
X-ray and gamma-ray fluxes, and Doppler shift in source f1 as a function of time.
 }
\label{fig3}
\end{figure}

\begin{figure}
\centerline{\includegraphics[scale=0.6]{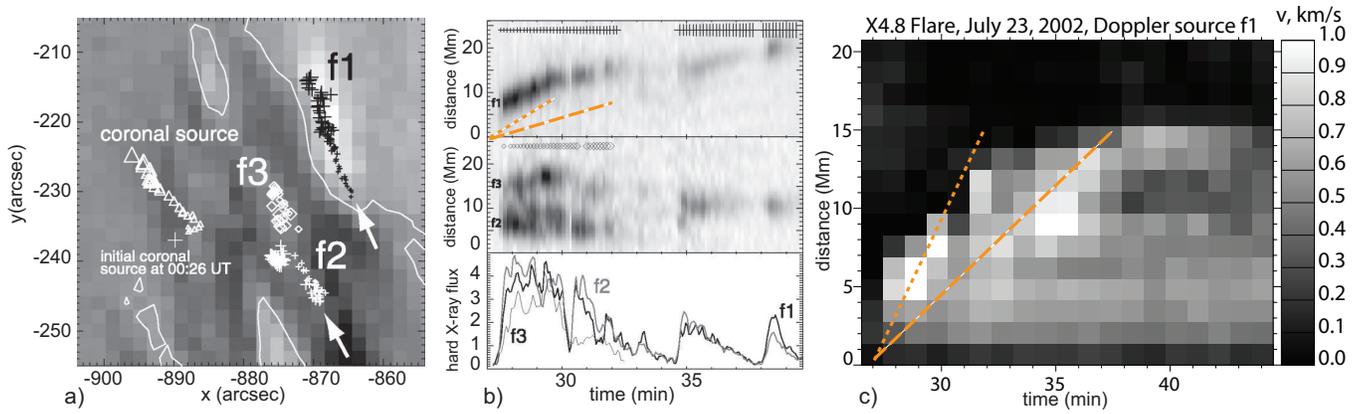}} \caption{a)
Evolution of HRX sources. The increasing size of the symbols represents
times from 00:26:35 to 00:39:07 UT, and b) HXR profiles (black in the top two panels
shows enhanced emission) along the ribbons, showing motion
with speed of up to 50 km/s (from \citet{Krucker2003}). c)
Doppler velocity profiles along the f1 ribbon, showing motion with
averaged speed ~25 km/s. The inclined orange lines correspond to 25 km/s (long dash)
and 50 km/s (short dash).
 }
\label{fig4}
\end{figure}

\begin{figure}
\centerline{\includegraphics[scale=0.5]{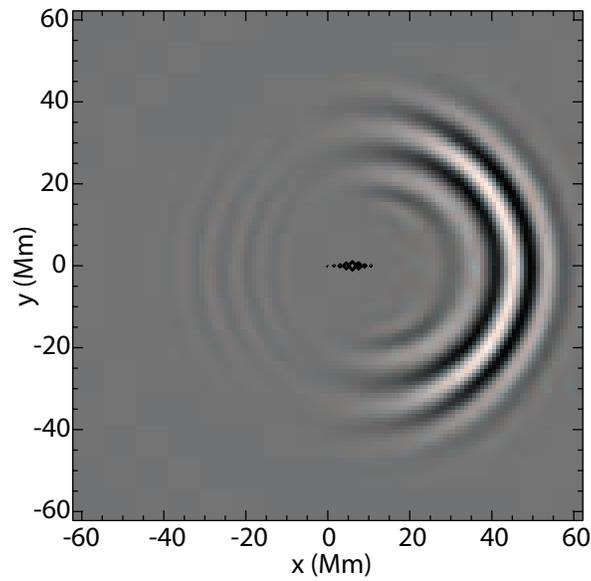}}
\caption{Theoretical model of seismic waves from a moving source,
which explains the observed anisotropy of sunquakes. The point
impulsive source is moving in the $x$ direction with the constant
speed of 25 km/s. Its strength as a function of time has a Gaussian
shape with FWHM of 3 min. The locations of the source are
shown by black diamonds at the center, the size of which is
proportional to the source strength.
 }
\label{fig5}
\end{figure}

The general picture that comes from the analysis of
MDI and RHESSI data is consistent with the previously developed
hydrodynamic thick-target model, illustrated in Fig.~\ref{fig2}
\citep{Kostiuk1975, Livshits1981, Fisher1985, Kosovichev1986}.  In
this model, high-energy electrons heat the upper chromosphere to high temperatures
generating a high-pressure region, expansion of which causes
evaporation of the chromospheric plasma and a high compression shock.
The shock reaches the photosphere and excites the seismic waves.
However, the new results show that it is important to include
effects of the multiple impact and moving source in the thick-target and
sunquake models.

The photospheric and helioseismic effects observed during the impulsive
phase of solar flares are closely related to the processes of acceleration
and propagation of electrons and ions, and may provide new important
information about these processes.

\clearpage

\clearpage


\clearpage

\begin{thebibliography}{}

\bibitem[Ben-Menahem(1962)]{Ben-Menahem1962} Ben-Menahem, A.\ 1962,
\jgr, 67, 345


\bibitem[Donea et al.(1999)]{Donea1999} Donea, A.-C., Braun, D. C. {\&} Lindsey, C., 1999,
\apj, 513, L143

\bibitem[Donea \& Lindsey(2005)]{Donea2005} Donea, A.-C. {\&} Lindsey, C., 2005,
\apj, 630, 1168

\bibitem[Fisher et al(1985)]{Fisher1985} Fisher, G. H., Canfield, R. C. \& McClymont, A. N., 1985, \apj, 289, 434

\bibitem[Hurford et al.(2003)]{Hurford2003} Hurford, G.~J.,
Schwartz, R.~A., Krucker, S., Lin, R.~P., Smith, D.~M., \& Vilmer,
N.\ 2003, \apjl, 595, L77


\bibitem[Hurford et al.(2006)]{Hurford2006} Hurford, G.~J.,
Krucker, S., Lin, R.~P., Schwartz, R.~A., Share, G.~H., \& Smith, D.~M.\
2006, \apjl, 644, L93

\bibitem[Kosovichev(1986)]{Kosovichev1986} Kosovichev, A.G., 1986,
Bull. Crimean Astrophys. Obs., 75,  6

\bibitem[Kosovichev \& Zharkova(1995)]{Kosovichev1995} Kosovichev, A. G.
{\&} Zharkova, V. V., 1995,
 Seismic Response to Solar Flares: Theoretical Predictions,
 in Helioseismology. ESA SP, Proc. 4th Soho
Workshop, p.341

\bibitem[Kosovichev \& Zharkova(1998)]{Kosovichev1998} Kosovichev, A. G. {\&} Zharkova, V. V., 1998,
Nature, 393, 317


\bibitem[Kosovichev(2006a)]{Kosovichev2006a} Kosovichev, A.~G.\ 2006a,
\solphys, 238, 1


\bibitem[Kosovichev(2006b)]{Kosovichev2006b} Kosovichev, A.~G.\ 2006b,
Direct Observations of Acoustic Waves Excited by Solar Flares and
their Propagation in Sunspot Regions, in: Solar MHD Theory and
Observations: A High Spatial Resolution Perspective, ASP Conference
Series, Vol. 354, p.154

\bibitem[Kosovichev(2006c)]{Kosovichev2006c} Kosovichev, A.~G.\ 2006c,
Sunquake sources and wave propagation, in: Proceedings of SOHO
18/GONG 2006/HELAS I, Beyond the spherical Sun, ESA SP-624, p.134.1



\bibitem[Kostiuk \& Pikelner(1975)]{Kostiuk1975} Kostiuk, N. D.
{\&} Pikelner, S. B., 1975, Sov. Astr., 18, 590

\bibitem[Krucker et al.(2003)]{Krucker2003} Krucker, S., Hurford,
G.~J., \& Lin, R.~P.\ 2003, \apjl, 595, L103

\bibitem[Lin et al.(2002)]{Lin2002} Lin, R.~P., et al.\ 2002,
\solphys, 210, 3



\bibitem[Livshits et al.(1981)]{Livshits1981} Livshits, M. A., Badalian, O. G.,
Kosovichev, A. G. {\&} Katsova, M. M., 1981, Sol. Phys., 73, 269

\bibitem[Medrek et al.(2000)]{Medrek2000} Medrek, M., Murawski,
K., \& Nakariakov, V.\ 2000, Acta Astronomica, 50, 405

\bibitem[Murphy et al.(2003)]{Murphy2003} Murphy, R.~J., Share,
G.~H., Hua, X.-M., Lin, R.~P., Smith, D.~M., \& Schwartz, R.~A.\ 2003,
\apjl, 595, L93



\bibitem[Podesta(2005)]{Podesta2005} Podesta, J.~J.\ 2005,
\solphys, 232, 1

\bibitem[Scherrer et al.(1995)]{Scherrer1995} Scherrer, P.~H., et
al.\ 1995, \solphys, 162, 129



\bibitem[Zharkova \& Zharkov(2007)]{Zharkova2007} Zharkova, V.~V.,
\& Zharkov, S.~I.\ 2007, \apj, 664, 573

\end{thebibliography}
\end{document}